\documentclass[prl,twocolumn,superscriptaddress,showpacs]{revtex4}

\usepackage{graphicx}
\usepackage{bm}

\newcommand{\beq}{\begin{equation}}
\newcommand{\eeq}{\end{equation}}
\newcommand{\beqa}{\begin{eqnarray}}
\newcommand{\eeqa}{\end{eqnarray}}

\begin{document}

\title{Measuring errors in single qubit rotations by pulsed electron
  paramagnetic resonance}

\author{John~J.~L.~Morton}
\email{john.morton@materials.ox.ac.uk} \affiliation{Department of
Materials, Oxford University, Oxford OX1 3PH, United Kingdom}

\author{Alexei~M.~Tyryshkin}
\affiliation{Department of Electrical Engineering, Princeton
University, Princeton, NJ 08544, USA}

\author{Arzhang~Ardavan}
\affiliation{Department of Materials, Oxford University, Oxford
OX1 3PH, United Kingdom}
\affiliation{Clarendon Laboratory,
Department of Physics, Oxford University, Oxford OX1 3PU, United
Kingdom}

\author{Kyriakos~Porfyrakis}
\affiliation{Department of Materials, Oxford University, Oxford
OX1 3PH, United Kingdom}

\author{S.~A.~Lyon}
\affiliation{Department of Electrical Engineering, Princeton
University, Princeton, NJ 08544, USA}

\author{G.~Andrew~D.~Briggs}
\affiliation{Department of Materials, Oxford University, Oxford
OX1 3PH, United Kingdom}

\date{\today}

\begin{abstract}
The ability to measure and reduce systematic errors in
single-qubit logic gates is crucial when evaluating quantum
computing implementations. We describe pulsed electron
paramagnetic resonance (EPR) sequences that can be used to measure
precisely even small systematic errors in rotations of
electron-spin-based qubits. Using these sequences we obtain values
for errors in rotation angle and axis for single-qubit rotations
using a commercial EPR spectrometer.  We conclude that errors in
qubit operations by pulsed EPR are not limiting factors in the
implementation of electron-spin based quantum computers.
\end{abstract}


\maketitle


Pulsed magnetic resonance methods have provided a useful
playground in which to test different aspects of quantum
computation. Single-qubit operations (spin rotations) can be
conveniently performed using classical radio frequency (RF) pulses
whilst two-qubit operations can be naturally realized through
exchange or dipolar interactions. Exploiting these advantages, the
largest-scale quantum computations to date have been demonstrated
using nuclear magnetic resonance (NMR) in liquid
solution~\cite{nielsen00,firstNMRQC}. Electron paramagnetic
resonance (EPR) offers many parallels with NMR, along with the key
advantage that the electron gyromagnetic ratio is of the order of
a thousand times larger. Thus pure ground states are
experimentally accessible in EPR, avoiding the scalability issues
surrounding NMR implementations~\cite{warren1997}.

For this reason, EPR has become a key element in several solid
state quantum information processing (QIP)
proposals~\cite{kane,burkard00,harneit,briggsRS,lyon}. The
practicality of these proposals is critically dependent on the
errors that are inherent in pulsed magnetic resonance experiments.
In particular, while the decoherence time ($T_2$) is generally
quoted as the ultimate figure of merit for qubit implementations,
understanding and minimizing the systematic errors inherent in
qubit manipulations is equally important. A number of general
approaches to tackling different classes of systematic error in
qubit rotations, employing composite rotation sequences, have been
proposed~\cite{levitt1986,jones03}. However, before these
approaches can be exploited practically, it is necessary to
characterize (and reduce) the errors that are associated with a
single rotation pulse.

There are two principal types of systematic error associated with
a single-qubit rotation: rotation angle error and rotation axis
error. In magnetic resonance experiments, rotation angle errors
arise from an uncertainty in the Rabi oscillation period,
associated with uncertainty in the magnitude and duration of the
applied RF pulse. On the other hand, rotation axis errors arise
from uncertainty in direction of the RF magnetic field in the
transverse plane of the rotating frame.

In this Letter, we show how multi-pulse sequences can be applied
to measure precisely these two classes of error.  These sequences
include some originally developed for NMR, as well as a novel
sequence developed specifically for quantifying phase errors. In
each case the sequence amplifies the errors, so that even small
errors are detectable.  We find that when comparing a commercial
EPR spectrometer to an NMR system which has been optimised for
high-fidelity qubit operations, rotation angle errors are as good,
whilst phase errors in EPR are worse by as much as an order of
magnitude.

The paramagnetic species used here to perform error measurements
in pulsed EPR is \emph{i}-NC$_{60}$ (also known as $N@C_{60}$),
consisting of an isolated nitrogen atom in the $^4$S$_{3/2}$
electronic state incarcerated by a C$_{60}$ fullerene cage. It is
an ideal system for these measurements because of its extremely
narrow EPR linewidth and long relaxation time in liquid
solution~\cite{Dietel99,Knapp1997}. $T_{2}$ has been measured to
be $80~\mu$s at room temperature, rising to $240~\mu$s at
170~K~\cite{tyryshkin03}.

The production and subsequent purification of \emph{i}-NC$_{60}$
is described elsewhere~\cite{mito}. High-purity \emph{i}-NC$_{60}$
powder was dissolved in CS$_{2}$ to a final concentration of
10$^{15}$/cm$^3$, freeze-pumped in three cycles to remove oxygen,
and finally sealed in a quartz EPR tube. Samples were 0.7-1.4~cm
long, and contained approximately $5\cdot 10^{13}$
\emph{i}-NC$_{60}$ spins. Pulsed EPR measurements were done at
190~K using an X-band Bruker Elexsys580e spectrometer, equipped
with a nitrogen-flow cryostat (Janis Research).

\emph{i}-NC$_{60}$ has electron spin $S=3/2$ coupled to the
$^{14}$N nuclear spin $I=1$. The EPR spectrum consists of three
lines centered at electron g-factor $g=2.003$ and split by
$^{14}$N hyperfine interaction $a=0.56$~mT in
CS$_2$~\cite{Murphy1996}. However, all pulsed EPR experiments
discussed below were done using the center line in the EPR
triplet, corresponding to the $^{14}$N nuclear spin projection
$M_I=0$. Given the small \emph{isotropic} hyperfine coupling in
\emph{i}-NC$_{60}$, the transitions with simultaneous flip of
electron and nuclear spins are largely forbidden and therefore the
evolution of electron spin can be treated individually for each
nuclear spin manifold. The Bloch sphere is a useful aid to
visualise the action of the pulse sequences described
below~\cite{schweiger}. While this model maps conveniently the
magnetization evolution of a single $S$ spin system, the evolution
of a coupled $S=3/2, I=1$ system is harder to visualize. It can be
shown that the evolution of the electron spin in the $M_I=0$
manifold is largely unaffected by the presence of the hyperfine
coupling and is adequately described by the classical vector
model~\cite{tyryshkin03}.


In an EPR experiment, a qubit rotation is achieved by applying an
on-resonance microwave pulse of controlled power and duration.
With a pulsed magnetic field strength $B_1$ and a pulse duration
$t$ the rotation angle is:
\begin{equation}\label{rotangle}
\theta=g\mu _{B}B_{1}t/\hbar .
\end{equation}
Possible off-resonance effects are neglected here because the EPR
linewidth for \emph{i}-NC$_{60}$ is much smaller than the excitation
width of the RF pulses used in our experiments.  According to
Eqn.~\ref{rotangle}, rotation angle errors may arise from either pulse
duration errors (which can be assumed uniform throughout the sample),
or errors in the magnitude of $B_{1}$ (which can vary across the
sample depending on the homogeneity of the EPR cavity mode).  The
hardware of the EPR spectrometer limits pulse duration to a 2~ns
resolution
and pulse power to 0.1~dB resolution. These together can
contribute of the order of 0.5$\%$ to the rotation angle error for
a typical 32~ns pulse.

\begin{figure}[t]
\centerline {\includegraphics[width=2.9in]{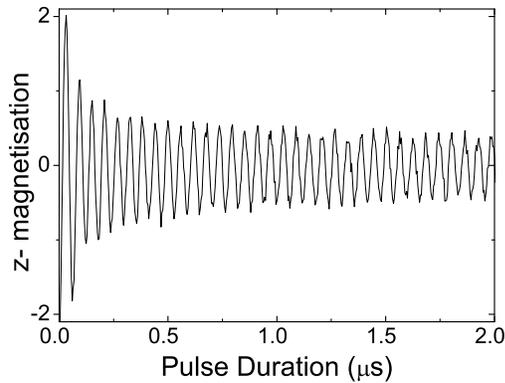}}
\caption{Rabi oscillations for \emph{i}-NC$_{60}$ in CS$_2$ at 190
K. } \label{LongPulse}
\end{figure}

In EPR, measurements are made of the magnetisation of the ensemble
in the $x - y$ plane, for a static magnetic field applied along
$z$. Hence, a simple measurement of the rotation angle error could
be made after a nominal $\pi/2$ rotation, which in practice is
$\pi/2 + \delta$ where $\delta$ is the error. The measured signal
is proportional to $\cos(\delta)$.  However, this approach is
unsatisfactory for two reasons: for a good measurement, a
reference (an ideal rotation) is needed with which to compare the
imperfect rotation; and the $\cos(\delta)$ term depends on
$\delta$ only to second order.

Another method involves applying a long RF pulse and observing
Rabi oscillations over a number of periods. The respective
experiment for \emph{i}-NC$_{60}$ is shown in
Fig.~\ref{LongPulse}, demonstrating Rabi oscillations whose
amplitudes decay at long pulse durations (over 80 oscillations
were seen). This decay is caused by inhomogeneity of $B_1$ fields
in the EPR cavity (spins are rotated with slightly different Rabi
frequencies and therefore gradually lose coherence), as well as
effects such as $B_0$ field inhomogeneity and the fact that the
output power
and phase
from the microwave amplifier
vary
at long pulse
lengths. In order to distill the rotation angle error from these
other effects it is necessary to use more sophisticated methods
for error measurement.

\begin{figure}[t]
\centerline {\includegraphics[width=3in]{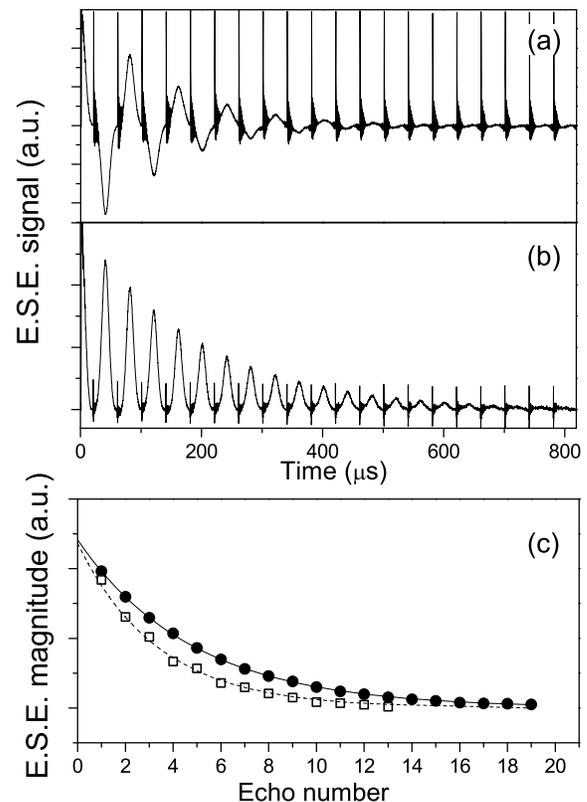}}
\caption{Comparison of the echo signal decays in (a) the CP and
(b) the CPMG pulse sequences. The narrow spikes correspond to the
applied $\pi$ pulses and the echo signals are seen in between. (c)
Decay of echo magnitudes for each sequence (CP: empty, CPMG:
filled). The solid grey line is a fit to a simple exponential
yielding $T_2=190$~$\mu$s; the dashed grey line is a fit yielding
rotation angle error parameters~\cite{cpeqn}.} \label{CPandCPMG}
\end{figure}

A better approach uses sequences of many pulses to compound the
error of a single pulse. A suitable sequence, CP
(Carr-Purcell~\cite{carrpurcell}), consists of a $\pi/2$ pulse
followed by a series of refocusing $\pi$ pulses, i.e. $\pi/2_{x} -
( \tau - \pi_{x} - \tau)_{n}$ in which pulse rotation angle
imperfections are additive~\cite{spinchoreo}. The typical
exponential decay of the echo is therefore further attenuated by
the cumulative rotation angle error, such that the echo amplitude
decays with a time constant shorter than $T_2$. Meiboom and Gill
proposed a modification of the CP sequence, termed
CPMG~\cite{meiboomgill}, which compensates for pulse length errors
by applying the refocussing $\pi$ pulses around the $y-$axis.


Figs.~\ref{CPandCPMG}(a) and (b) compare the trains of echo
signals observed in the CP and CPMG experiments for
\emph{i}-NC$_{60}$. The decay of the CPMG echo magnitudes is
fitted to a simple exponential to obtain the transverse relaxation
time $T_2$. In determining the decay of echo magnitudes due to
rotation angle errors in the CP sequence, we assume the flip angle
error follows a Gaussian distribution with mean $\delta_0$ (to
account for pulse duration errors) and standard deviation
$\sigma_{\delta}$ (to account for inhomogeneity in the oscillatory
magnetic field strength)~\cite{cpeqn}. From the fit we obtain
$\sigma_{\delta} = 18^{\circ}$ in every 180$^{\circ}$ rotation, or
approximately $10\%$.  This figure is consistent with the expected
inhomogeneity in the applied $B_{1}$ field, and was seen to vary
by altering the dimensions of the sample.

\begin{figure}[t]
\centerline{
\hspace{-1.5cm}\includegraphics[width=3in]{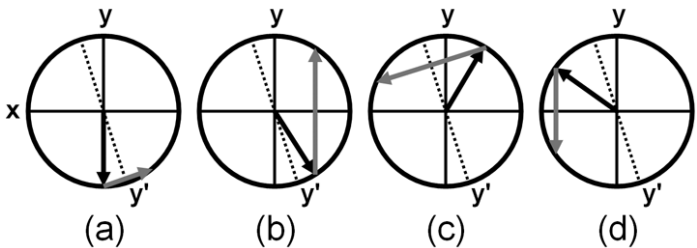}}
\caption{Evolution of the magnetisation vector in the rotating
frame during a SPAM experiment. The ideal $y$ and $x$ axes are
associated with in-phase and quadrature channels of the quadrature
detector, respectively. $\pi/2_x$ and $\pi_x$ pulses are assumed
to be along the $x$-axis; $\pi_y$ pulses deviate from the ideal
$y$-axis by the error angle $\delta$ and are oriented along $y'$.
The magnetisation vector (bold) is shown (a) immediately after the
$\pi/2_x$ pulse, and at times of echo formation after (b) the
first $\pi_y$ pulse, (c) the first $\pi_x$ pulse, (d) the second
$\pi_y$ pulse. Alternating the refocussing pulses between  $x$ and
$y'$ in SPAM results in an accumulation of error in the phase of
the echo signal. } \label{SPAMfig}
\end{figure}

Having determined the rotation angle errors, we now turn to
rotation axis errors. The Bruker spectrometer used in this work
offers four independent pulse-forming channels, each supplied with
uncalibrated analogue phase shifter. Our initial goal was to set
microwave phases in two pulse-forming channels to be orthogonal to
each other, which is to orient the $B_{1}$ field in one channel
along the $x$-axis and the other along the $y$-axis. The ideal $x$
and $y$-axes in the rotating frame are defined with respect to the
phase of quadrature detection channels, therefore, before
perfecting the phase setting of the pulse-forming channels we
first examined the orthogonality between the two channels of the
quadrature detector. This was done by applying a slightly
off-resonance $\pi/2$ pulse and observing the resulting FID
oscillations from each detection channel. A numerical fit revealed
an angle of $89.3\pm1^{\circ}$ between the nominally orthogonal
``real'' and ``imaginary'' detection channels.

Traditionally, the phase of a pulse-forming channel is adjusted by
applying a simple $\pi/2$ rotation about each channel and
observing the FID signal. The inherent imprecision of this
approach is not a serious problem in traditional EPR applications
in which only a few pulses are applied, but is potentially
devastating for the fidelity of a qubit state in a multiple-pulse
computation.

\begin{figure}[t]
\centerline {\includegraphics[width=2.8in]{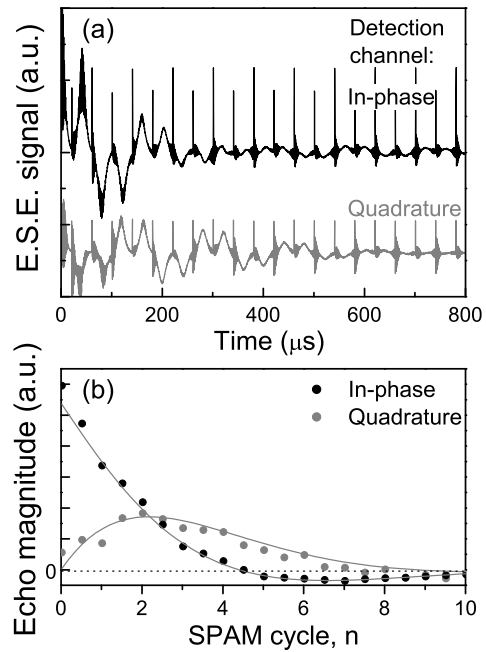}}
\caption{(a) The echo train from SPAM showing a leakage of signal
between the $y$ (in-phase: top trace) and $x$ (quadrature: bottom
trace) detection channels resulting from phase error between
nominal $\pi_x$ and $\pi_y$ pulses (traces are offset for
clarity). (b) Magnitudes of successive echoes from (a) fitted to
functions described in the text yield
$\delta=(10.3\pm0.5)^{\circ}$.} \label{SPAMdata}
\end{figure}

In order to measure the orthogonality of the rotation axes in two
pulse-forming channels with a high precision we designed a pulse
sequence to accumulate phase errors, by analogy with the way that
the CP sequence above accumulates flip angle error.  This
sequence, called Sequence for Phase-error AMplification (SPAM) is
given in Eqn.~\ref{SPAM} and illustrated in Fig.~\ref{SPAMfig}.
\begin{equation}\label{SPAM}
\pi/2_{x} - ( \tau - \pi_{y} - \tau - \tau - \pi_{x} - \tau)_{n}
\end{equation}
After $n$ cycles of this sequence with a phase error $\delta$
(i.e.\ the phase of the $\pi_y$ pulses is actually $\pi/2+\delta$
with respect to the $\pi_x$ pulses) the echo vector points along
$(\cos(2n\delta), \sin(2n\delta), 0)$. Thus, a non-zero $\delta$
in this sequence results in an accumulating ``leakage'' of the
echo amplitude into the orthogonal detection channel, with only a
second-order sensitivity to the pulse length errors described
above.

Fig.~\ref{SPAMdata}(a) shows the echo train obtained in a SPAM
sequence for \emph{i}-NC$_{60}$. For this experiment a phase error
of approximately 10$^{\circ}$ between the $\pi_x$ and $\pi_y$
pulses was intentionally introduced by observing the FID signals.
The measured echo magnitudes for the real and imaginary channels
are described by $\cos \left(2n\delta\right)$ and
$\sin\left(2n\delta\right)$ respectively, with a exponential decay
envelope $\exp\left(-{4n\tau/T_{2}}\right)$. $n$ is the number of
SPAM cycles (each comprising two refocusing pulses) and $T_{2}$ is
taken from a CPMG experiment described above.
Fig.~\ref{SPAMdata}(b) shows the echo amplitudes extracted from
(a). The fit yields $\delta = (10.3\pm0.5)^{\circ}$, which is
close to the intended phase error. This demonstrates that SPAM is
a useful way of both measuring the phase error in nominally
orthogonal channels and accurately setting arbitrary phases
between channels.

Orthogonality of the $\pi_x$ and $\pi_y$ channels was then
optimised by traditional procedures (by observing the FID as
described above), and a phase error of $\delta =
(1.5\pm0.3)^{\circ}$ was measured from a SPAM sequence. This
provides a measure of the phase error typical in conventional
pulsed EPR experiments. The fact that it can be measured to a
precision of about $0.3^\circ$ using a SPAM sequence allows us to
reduce it substantially further. Careful optimisation of the phase
error using the SPAM sequence yielded $\delta =
(0.3\pm0.1)^{\circ}$.


In summary, we find that although commercial pulsed EPR
spectrometers have not been designed with high precision
operations and multi-pulse sequences in mind, the Bruker machine
performs very well. The rotation angle error of about 10\% is
almost entirely due to inhomogeneity of the oscillatory magnetic
field, and is comparable with those typically observed in
NMR~\cite{jones03}.
The rotation axis
(phase) error of about 0.3$^{\circ}$ is also comparable with those
typically encountered in NMR (though worse than in optimised
quantum computing NMR spectrometers).

The most significant strength of the SPAM methodology is that it
provides a method of setting relative phases between channels with
very high precision. This makes it possible to exploit techniques
developed within the context of NMR for applying sequences of
pulses at various phases that correct for rotation angle
errors~\cite{jones03, wimperis, chuang04}. Such a sequence can
straightforwardly reduce a rotation angle error of order
$\epsilon$ to $\epsilon^6$. Using this approach, our results imply
that rotation angle errors can be reduced the order of $10^{-6}$,
well within the threshold of $10^{-4}$ often cited for
fault-tolerant quantum computation~\cite{steane}. This is the
subject of a subsequent paper.

We have shown that the technology exists to transfer the quantum
information processing methodology that has been developed in the
context of NMR to EPR, thereby overcoming the scaling limitations
associated with NMR. We conclude that errors in qubit operations
do not restrict the viability of an EPR-based quantum computer.
Furthermore, we demonstrate the successful application of long
pulse sequences (necessary for running quantum algorithms) to an
EPR qubit candidate, the \emph{i}-NC$_{60}$ molecule. Finally, we
have demonstrated a set of pulse sequences that can be used to
amplify and measure precisely the effect of rotation angle and
axis errors in any NMR or EPR pulsed magnetic resonance
spectrometer.

We would like to thank Wolfgang Harneit's group at the
Hahn-Meitner Institute for providing Nitrogen-doped fullerenes,
and John Dennis at Queen Mary's College, London, Martin Austwick
and Gavin Morley for the purification of \emph{i}-NC$_{60}$. We
also thank Jonathan Jones for stimulating and valuable
discussions. A Foresight LINK grant \emph{Nanoelectronics at the
quantum edge}, an EPSRC grant and the Oxford-Princeton Link fund
supported this project. AA is supported by the Royal Society. Work
at Princeton was supported by the NSF International Office through
the Princeton MRSEC Grant No. DMR-0213706 and by the ARO and ARDA
under Contract No. DAAD19-02-1-0040.


\clearpage
\newpage
\setcounter{equation}{0}
\appendix
\section{Supporting Material}

In the Carr-Purcell (CP) sequence, $\pi/2_{x} - ( \tau - \pi_{x} -
\tau)_{n}$, the effect of rotation angle errors is complicated by
the dispersion of spins in the $x-y$ plane due to $B_0$ field
inhomogeneity. For example, those spins pointing along $y$ when
the refocussing $\pi_x$ pulse is applied pick up the most error,
whilst those pointing along $x$ are unaffected.  We assume a
uniform distribution of phase, i.e. $\tau >> T_2^*$, and a
Gaussian distribution of rotation angle error with mean $\delta_0$
and standard deviation $\sigma$.  The echo magnitudes obey the
following equation, after the $n$th pulse in the sequence:

\beq A_{CP}(n)=1-\sum_{m=1}^n\left(a_m + \sum_{k=1}^m b_k
\exp{\left(\frac{-\sigma^2k^2}{2}\right)}\cos{\left(k\delta_0\right)}\right)\eeq

\beq a_m=\frac{~^{2m}C_m ~^{n+m-1}C_{2m-1}
~^{1/2}C_{m}~n~(2m-1)}{2m} \eeq

\beq b_k=\frac{(-1)^k ~^{2m}C_{m-k} ~^{n+m-1}C_{2m-1}
~^{1/2}C_{m}~n~(2m-1)}{m} \eeq\\
For $n\sigma<1$, and assuming $\delta_0=0$, this can be
approximated to Eqn. (4) below.

\beq A_{CP}(n)=\exp{\left(\frac{-\sigma^2n^2}{4}\right)} \eeq

To fit the CP decay we use $A_{CP}(n) \exp{(-t/T_2)}$, with $T_2$
taken from the CPMG fit.

\clearpage


\begin{thebibliography}{23}
\expandafter\ifx\csname
natexlab\endcsname\relax\def\natexlab#1{#1}\fi
\expandafter\ifx\csname bibnamefont\endcsname\relax
  \def\bibnamefont#1{#1}\fi
\expandafter\ifx\csname bibfnamefont\endcsname\relax
  \def\bibfnamefont#1{#1}\fi
\expandafter\ifx\csname citenamefont\endcsname\relax
  \def\citenamefont#1{#1}\fi
\expandafter\ifx\csname url\endcsname\relax
  \def\url#1{\texttt{#1}}\fi
\expandafter\ifx\csname
urlprefix\endcsname\relax\def\urlprefix{URL }\fi
\providecommand{\bibinfo}[2]{#2}
\providecommand{\eprint}[2][]{\url{#2}}

\bibitem[{\citenamefont{Nielsen and Chuang}(2000)}]{nielsen00}
\bibinfo{author}{\bibfnamefont{M.~A.} \bibnamefont{Nielsen}} \bibnamefont{and}
  \bibinfo{author}{\bibfnamefont{I.~L.} \bibnamefont{Chuang}},
  \emph{\bibinfo{title}{Quantum computation and quantum information}}
  (\bibinfo{publisher}{Cambridge University Press}, \bibinfo{address}{Cambridge
  ; New York}, \bibinfo{year}{2000}).

\bibitem[{\citenamefont{Cory et~al.}(1997)\citenamefont{Cory, Fahmy, and
  Havel}}]{firstNMRQC}
\bibinfo{author}{\bibfnamefont{D.~G.} \bibnamefont{Cory}},
  \bibinfo{author}{\bibfnamefont{A.~F.} \bibnamefont{Fahmy}}, \bibnamefont{and}
  \bibinfo{author}{\bibfnamefont{T.~F.} \bibnamefont{Havel}},
  \bibinfo{journal}{Proc. Natl. Acad. Sci. U. S. A.}
  \textbf{\bibinfo{volume}{94}}, \bibinfo{pages}{1634} (\bibinfo{year}{1997}).

\bibitem[{\citenamefont{Warren}(1997)}]{warren1997}
\bibinfo{author}{\bibfnamefont{W.~S.} \bibnamefont{Warren}},
  \bibinfo{journal}{Science} \textbf{\bibinfo{volume}{277}},
  \bibinfo{pages}{1688} (\bibinfo{year}{1997}).

\bibitem[{\citenamefont{Kane}(1998)}]{kane}
\bibinfo{author}{\bibfnamefont{B.~E.} \bibnamefont{Kane}},
  \bibinfo{journal}{Nature} \textbf{\bibinfo{volume}{393}},
  \bibinfo{pages}{133} (\bibinfo{year}{1998}).

\bibitem[{\citenamefont{Burkard et~al.}(2000)\citenamefont{Burkard, Engel, and
  Loss}}]{burkard00}
\bibinfo{author}{\bibfnamefont{G.}~\bibnamefont{Burkard}},
  \bibinfo{author}{\bibfnamefont{H.~A.} \bibnamefont{Engel}}, \bibnamefont{and}
  \bibinfo{author}{\bibfnamefont{D.}~\bibnamefont{Loss}},
  \bibinfo{journal}{Fortschritte Phys.-Prog. Phys.}
  \textbf{\bibinfo{volume}{48}}, \bibinfo{pages}{965} (\bibinfo{year}{2000}).

\bibitem[{\citenamefont{Harneit}(2002)}]{harneit}
\bibinfo{author}{\bibfnamefont{W.}~\bibnamefont{Harneit}},
  \bibinfo{journal}{Phys. Rev. A} \textbf{\bibinfo{volume}{65}},
  \bibinfo{pages}{032322} (\bibinfo{year}{2002}).

\bibitem[{\citenamefont{Ardavan et~al.}(2003)\citenamefont{Ardavan, Austwick,
  Benjamin, Briggs, Dennis, Ferguson, Hasko, Kanai, Khlobystov, Lovett
  et~al.}}]{briggsRS}
\bibinfo{author}{\bibfnamefont{A.}~\bibnamefont{Ardavan}},
  \bibinfo{author}{\bibfnamefont{M.}~\bibnamefont{Austwick}},
  \bibinfo{author}{\bibfnamefont{S.~C.} \bibnamefont{Benjamin}},
  \bibinfo{author}{\bibfnamefont{G.~A.~D.} \bibnamefont{Briggs}},
  \bibinfo{author}{\bibfnamefont{T.~J.~S.} \bibnamefont{Dennis}},
  \bibinfo{author}{\bibfnamefont{A.}~\bibnamefont{Ferguson}},
  \bibinfo{author}{\bibfnamefont{D.~G.} \bibnamefont{Hasko}},
  \bibinfo{author}{\bibfnamefont{M.}~\bibnamefont{Kanai}},
  \bibinfo{author}{\bibfnamefont{A.~N.} \bibnamefont{Khlobystov}},
  \bibinfo{author}{\bibfnamefont{B.~W.} \bibnamefont{Lovett}},
  \bibnamefont{et~al.}, \bibinfo{journal}{Philos. Trans. R. Soc. Lond. Ser.
  A-Math. Phys. Eng. Sci.} \textbf{\bibinfo{volume}{361}},
  \bibinfo{pages}{1473} (\bibinfo{year}{2003}).

\bibitem[{\citenamefont{Lyon}(http://arXiv.org/abs/cond-mat/0301581)}]{lyon}
\bibinfo{author}{\bibfnamefont{S.~A.} \bibnamefont{Lyon}}
  (\bibinfo{year}{http://arXiv.org/abs/cond-mat/0301581}).

\bibitem[{\citenamefont{Levitt}(1986)}]{levitt1986}
\bibinfo{author}{\bibfnamefont{M.~H.} \bibnamefont{Levitt}},
  \bibinfo{journal}{Prog. Nucl. Magn. Reson. Spectrosc.}
  \textbf{\bibinfo{volume}{18}}, \bibinfo{pages}{61} (\bibinfo{year}{1986}).

\bibitem[{\citenamefont{Cummins et~al.}(2003)\citenamefont{Cummins, Llewellyn,
  and Jones}}]{jones03}
\bibinfo{author}{\bibfnamefont{H.~K.} \bibnamefont{Cummins}},
  \bibinfo{author}{\bibfnamefont{G.}~\bibnamefont{Llewellyn}},
  \bibnamefont{and} \bibinfo{author}{\bibfnamefont{J.~A.} \bibnamefont{Jones}},
  \bibinfo{journal}{Phys. Rev. A} \textbf{\bibinfo{volume}{67}},
  \bibinfo{pages}{042308} (\bibinfo{year}{2003}).

\bibitem[{\citenamefont{Dietel et~al.}(1999)\citenamefont{Dietel, Hirsch,
  Pietzak, Waiblinger, Lips, Weidinger, Gruss, and Dinse}}]{Dietel99}
\bibinfo{author}{\bibfnamefont{E.}~\bibnamefont{Dietel}},
  \bibinfo{author}{\bibfnamefont{A.}~\bibnamefont{Hirsch}},
  \bibinfo{author}{\bibfnamefont{B.}~\bibnamefont{Pietzak}},
  \bibinfo{author}{\bibfnamefont{M.}~\bibnamefont{Waiblinger}},
  \bibinfo{author}{\bibfnamefont{K.}~\bibnamefont{Lips}},
  \bibinfo{author}{\bibfnamefont{A.}~\bibnamefont{Weidinger}},
  \bibinfo{author}{\bibfnamefont{A.}~\bibnamefont{Gruss}}, \bibnamefont{and}
  \bibinfo{author}{\bibfnamefont{K.~P.} \bibnamefont{Dinse}},
  \bibinfo{journal}{J. Am. Chem. Soc.} \textbf{\bibinfo{volume}{121}},
  \bibinfo{pages}{2432} (\bibinfo{year}{1999}).

\bibitem[{\citenamefont{Knapp et~al.}(1997)\citenamefont{Knapp, Dinse, Pietzak,
  Waiblinger, and Weidinger}}]{Knapp1997}
\bibinfo{author}{\bibfnamefont{C.}~\bibnamefont{Knapp}},
  \bibinfo{author}{\bibfnamefont{K.~P.} \bibnamefont{Dinse}},
  \bibinfo{author}{\bibfnamefont{B.}~\bibnamefont{Pietzak}},
  \bibinfo{author}{\bibfnamefont{M.}~\bibnamefont{Waiblinger}},
  \bibnamefont{and}
  \bibinfo{author}{\bibfnamefont{A.}~\bibnamefont{Weidinger}},
  \bibinfo{journal}{Chem. Phys. Lett.} \textbf{\bibinfo{volume}{272}},
  \bibinfo{pages}{433} (\bibinfo{year}{1997}).

\bibitem[{\citenamefont{Tyryshkin et~al.}()\citenamefont{Tyryshkin, Morton,
  Ardavan, Porfyrakis, Lyon, and Briggs}}]{tyryshkin03}
\bibinfo{author}{\bibfnamefont{A.~M.} \bibnamefont{Tyryshkin}},
  \bibinfo{author}{\bibfnamefont{J.~J.~L.} \bibnamefont{Morton}},
  \bibinfo{author}{\bibfnamefont{A.}~\bibnamefont{Ardavan}},
  \bibinfo{author}{\bibfnamefont{K.}~\bibnamefont{Porfyrakis}},
  \bibinfo{author}{\bibfnamefont{S.~A.} \bibnamefont{Lyon}}, \bibnamefont{and}
  \bibinfo{author}{\bibfnamefont{G.~A.~D.} \bibnamefont{Briggs}},
  \eprint{\emph{in preparation}}.

\bibitem[{\citenamefont{Kanai et~al.}()\citenamefont{Kanai, Porfyrakis, Briggs,
  and Dennis}}]{mito}
\bibinfo{author}{\bibfnamefont{M.}~\bibnamefont{Kanai}},
  \bibinfo{author}{\bibfnamefont{K.}~\bibnamefont{Porfyrakis}},
  \bibinfo{author}{\bibfnamefont{G.~A.~D.} \bibnamefont{Briggs}},
  \bibnamefont{and} \bibinfo{author}{\bibfnamefont{T.~J.~S.}
  \bibnamefont{Dennis}}, \eprint{Chem. Commun. \emph{in press.}}

\bibitem[{\citenamefont{Almeida-Murphy
  et~al.}(1996)\citenamefont{Almeida-Murphy, Pawlik, Weidinger, Hohne, Alcala,
  and Spaeth}}]{Murphy1996}
\bibinfo{author}{\bibfnamefont{T.}~\bibnamefont{Almeida-Murphy}},
  \bibinfo{author}{\bibfnamefont{T.}~\bibnamefont{Pawlik}},
  \bibinfo{author}{\bibfnamefont{A.}~\bibnamefont{Weidinger}},
  \bibinfo{author}{\bibfnamefont{M.}~\bibnamefont{Hohne}},
  \bibinfo{author}{\bibfnamefont{R.}~\bibnamefont{Alcala}}, \bibnamefont{and}
  \bibinfo{author}{\bibfnamefont{J.~M.} \bibnamefont{Spaeth}},
  \bibinfo{journal}{Phys. Rev. Lett.} \textbf{\bibinfo{volume}{77}},
  \bibinfo{pages}{1075} (\bibinfo{year}{1996}).

\bibitem[{\citenamefont{Schweiger and Jeschke}(2001)}]{schweiger}
\bibinfo{author}{\bibfnamefont{A.}~\bibnamefont{Schweiger}} \bibnamefont{and}
  \bibinfo{author}{\bibfnamefont{G.}~\bibnamefont{Jeschke}},
  \emph{\bibinfo{title}{Principles of pulse electron paramagnetic resonance}}
  (\bibinfo{publisher}{Oxford University Press}, \bibinfo{address}{New York},
  \bibinfo{year}{2001}).

\bibitem[{cpe()}]{cpeqn}
\eprint{See supporting material}.

\bibitem[{\citenamefont{Carr and Purcell}(1954)}]{carrpurcell}
\bibinfo{author}{\bibfnamefont{H.~Y.} \bibnamefont{Carr}} \bibnamefont{and}
  \bibinfo{author}{\bibfnamefont{E.~M.} \bibnamefont{Purcell}},
  \textbf{\bibinfo{volume}{94}}, \bibinfo{pages}{630} (\bibinfo{year}{1954}).

\bibitem[{\citenamefont{Freeman}(1997)}]{spinchoreo}
\bibinfo{author}{\bibfnamefont{R.}~\bibnamefont{Freeman}},
  \emph{\bibinfo{title}{Spin choreography : basic steps in high resolution
  NMR}} (\bibinfo{publisher}{Spektrum ; University Science Books},
  \bibinfo{address}{Oxford Sausalito, Calif.}, \bibinfo{year}{1997}).

\bibitem[{\citenamefont{Meiboom and Gill}(1958)}]{meiboomgill}
\bibinfo{author}{\bibfnamefont{S.}~\bibnamefont{Meiboom}} \bibnamefont{and}
  \bibinfo{author}{\bibfnamefont{D.}~\bibnamefont{Gill}},
  \bibinfo{journal}{Rev. Sci. Instrum.} \textbf{\bibinfo{volume}{29}},
  \bibinfo{pages}{688} (\bibinfo{year}{1958}).

\bibitem[{\citenamefont{Wimperis}(1994)}]{wimperis}
\bibinfo{author}{\bibfnamefont{S.}~\bibnamefont{Wimperis}},
  \bibinfo{journal}{Journal of Magnetic Resonance}
  \textbf{\bibinfo{volume}{109}}, \bibinfo{pages}{221} (\bibinfo{year}{1994}).

\bibitem[{\citenamefont{Brown et~al.}()\citenamefont{Brown, Harrow, and
  Chuang}}]{chuang04}
\bibinfo{author}{\bibfnamefont{K.}~\bibnamefont{Brown}},
  \bibinfo{author}{\bibfnamefont{A.}~\bibnamefont{Harrow}}, \bibnamefont{and}
  \bibinfo{author}{\bibfnamefont{I.}~\bibnamefont{Chuang}},
  \eprint{quant-ph/0407022}.

\bibitem[{\citenamefont{Steane}(2003)}]{steane}
\bibinfo{author}{\bibfnamefont{A.~M.} \bibnamefont{Steane}},
  \bibinfo{journal}{Physical Review A} \textbf{\bibinfo{volume}{68}},
  \bibinfo{pages}{042322} (\bibinfo{year}{2003}).

\end{thebibliography}
\end{document}